\newcounter{myctr}
\def\myitem{\refstepcounter{myctr}\bibfont\noindent\ifnum\themyctr>9\else\phantom{0}\fi\hangindent17pt\themyctr.\enskip}

\documentclass{ws-ijqi}

\begin{document}

\markboth{M. Scala {\em et al.}} {Non-Markovian dynamics of cavity
losses}

%%%%%%%%%%%%%%%%%%%%% Publisher's Area please ignore %%%%%%%%%%%%%%
\catchline{}{}{}{}{}
%%%%%%%%%%%%%%%%%%%%%%%%%%%%%%%%%%%%%%%%%%%%%%%%%%%%%%%%%%%%%%%%%%%

\title{NON-MARKOVIAN DYNAMICS OF CAVITY LOSSES}

\author{MATTEO SCALA}

\address{Departamento de \'Optica, Facultad de F\'isica, Universidad
Complutense\\ Madrid, 28040, Spain\\
and Dipartimento di Scienze Fisiche ed Astronomiche,
Universit\`{a}
degli Studi di Palermo\\
Palermo, 90123, Italy
\\
matteo.scala@fisica.unipa.it}

\author{BENEDETTO MILITELLO, ANTONINO MESSINA}

\address{MIUR and Dipartimento di Scienze Fisiche ed Astronomiche,
Universit\`{a}
degli Studi di Palermo\\
Palermo, I-90123, Italy}

\author{JYRKI PIILO, SABRINA MANISCALCO, KALLE-ANTTI SUOMINEN}

\address{Department of Physics, University of Turku\\ Turun yliopisto,
FI-20014,
 Finland}

\maketitle

\begin{history}
\received{\today}
%\revised{Day Month Year}
%\accepted{Day Month Year}
%\comby{(xxxxxxxxxx)}
\end{history}

\begin{abstract}
We provide a microscopic derivation for the non-Markovian master
equation for an atom-cavity system with cavity losses and show
that they can induce population trapping in the atomic excited
state, when the environment outside the cavity has a non-flat
spectrum. Our results apply to hybrid solid state systems and can
turn out to be helpful to find the most appropriate description of
leakage in the recent developments of cavity quantum
electrodynamics.
\end{abstract}


\section{Introduction}
\label{sec_cqedintro} Cavity quantum electrodynamics (CQED) is one
of the most important fields of quantum optics, both from a
fundamental and an applicative point of view\cite{harochebook}.
Indeed, the interaction of an atom and a single mode of the
quantized electromagnetic field inside a high-$Q$ cavity can be
exploited to study the properties of highly non-classical states
of the atom-cavity system and to use their non-classical features
for quantum information processing\cite{harocheRMP}. In this
context it is fundamental to take into account the unavoidable
coupling between the atom-cavity system and the environment
external to it. In general, the coupling with an environment gives
rise to dissipation and decoherence
phenomena\cite{petruccionebook}, which can strongly affect the
dynamics of the quantum system under study. In previous
papers\cite{Scala,Scala2} we derived a master equation for an
atom-cavity system which takes into account from the very
beginning a coupling between the atom and the cavity described by
the Jaynes-Cummings (JC) model\cite{JC} . The approach followed in
Refs. \refcite{Scala} and \refcite{Scala2} is different from the
one usually reported in the literature to describe cavity losses.
Indeed the usual model of cavity losses has been introduced in a
phenomenological way, i.e., one microscopically derives the master
equation for the cavity only, described as a quantum harmonic
oscillator coupled to a bosonic bath, and then assumes that the
presence of the atom inside the cavity does not affect the
structure of the dissipator in the master equation\cite{haroche}.
Through the microscopic model derived in
Refs.~\refcite{Scala,Scala2} it is possible to give a complete
justification of the phenomenological model, which turns out to be
valid when the spectrum of the environment is flat and its
temperature is zero. The situation is quite different if the
environment has a non-flat spectral density. This is not an
uncommon situation nowadays: indeed there are new-generation CQED
experiments which are performed on high-Q cavities created inside
photonic bandgap (PBG) materials\cite{lambropoulosreview}. For
this system we proposed a non-Markovian model of cavity
losses\cite{Scala3} which predicts a dynamics completely different
from the one predicted by the phenomenological model, giving rise
to phenomena such as population trapping due to cavity losses.
Scope of this paper is to provide some details of the derivation
of the non-Markovian master equation for the JC model and to show
that the main predictions of the theory, namely population
trapping due to cavity losses\cite{Scala3}, does not depend on the
shape of the spectral density one chooses, but only on the values
of the asymptotic decay rates corresponding to the transitions of
interest. We clarify this point by taking into account two
different models of environmental spectral densities and showing
that, for the same asymptotic values of the decay rates, the
predictions differ only in the details of the very short-time
dynamics, while the two densities give the same predictions for
longer times. The paper is structured as follows. In Sec.~2 we
present the non-Markovian model of cavity losses, in Sec.~3 we
present the calculation of the time-dependent decay rate for a
Lorentzian spectrum and summarize the dynamics of the system in
this case. In Sec.~4 the case of a spectrum with a Lorentzian gap
is presented along with some conclusive remarks.

\section{The non-Markovian master equation for cavity
losses}\label{ch5_secMEq}

The system we study consists of a two-level atom interacting with
a mode of a cavity coupled to a bosonic environment. Calling
$\omega_0$ the Bohr frequency of the atom and $\left|g\right>$ and
$\left|e\right>$ its ground and its excited states respectively,
the interaction between the atom and the cavity mode is described,
at resonance and in units of $\hbar$, by the JC
Hamiltonian\cite{JC}
$H_{JC}=\frac{\,\omega_0}{2}\sigma_z+\omega_0\,a^\dag
 a+\Omega\left(a\sigma_++a^\dag\sigma_-\right)$,
where $a^\dag$ ($a$) is the creation (annihilation) operator of
the mode, $\sigma_-=\left|g\right>\left<e\right|$,
$\sigma_+=\left|e\right>\left<g\right|$, and $\sigma_z =
\left|e\rangle \langle e \right| - \left|g \rangle \langle g
\right|$.

The cavity mode interacts with a bosonic reservoir through an
interaction Hamiltonian linear in the bosonic operators of the
reservoir and of the cavity mode. More precisely we assume that
the dynamics of the total system (atom-cavity system and
environment) is described by a Hamiltonian $H=H_S+H_E+H_{\rm
int}$, where $H_S=H_{JC}$ is relative to the atom-cavity system,
$H_E=\sum_k\omega_k b_k^\dag b_k$ is relative to the environment
and $H_{\rm
int}=\left(a+a^\dag\right)\sum_kg_k\left(b_k+b_k^\dag\right)$
describes the system-environment interaction, with $\omega_k$ the
frequencies of the environment oscillators, $b^{\dag}_k$ ($b_k$)
the creation (annihilation) operator of quanta in the $k$-th
environmental mode, and $g_k$ the coupling constants.
%This model
%is equivalent to that used in Ref.~\cite{cohen_libro}, since the
%additional terms rapidly oscillating at frequencies of the order
%of $\omega_0$ are washed out in the  RWA performed in the
%following.
Since the reservoir causing cavity losses is immersed in the PBG
material, we expect its spectrum to be non-flat: then, to be
rigorous, the master equation must be derived in the framework of
a non-Markovian theory.

Using the second order of the time-convolutionless (TCL)
expansion\cite{petruccionebook} of an exact non-Markovian master
equation, and neglecting the atomic spontaneous emission and the
Lamb shifts, it is possible to show that for a weak enough damping
the master equation for the system is local in time and that its
structure, in the rotating wave approximation (RWA), is the same
of the Markovian master equation for the same system, with the
important difference that now the decay rates are time-dependent.
Examples of non-Markovian master equations of this type can be
found for example in Refs.~\refcite{petruccionebook},
\refcite{sabrina3}~and~\refcite{sabrina4}. In our analysis we will
focus on the case of one initial excitation only and we will
consider a reservoir at zero temperature. Then, the non-Markovian
master equation for the atom-cavity system density operator
$\rho$, in this case, is the following one\cite{Scala3}:
\begin{eqnarray}\label{masterequation}
 &&\dot{\rho}(t)=-i\left[H_{JC},\rho\right]\nonumber\\
\nonumber\\&&+\gamma\left(\omega_0+\Omega,t\right)
  \left(\frac{1}{2}\left|E_0\right>\left<E_{1,+}\right|\rho(t)\left|E_{1,+}\right>\left<E_0\right|
  -\frac{1}{4}\left\{\left|E_{1,+}\right>\left<E_{1,+}\right|,\rho(t)\right\}\right)\nonumber\\
\nonumber\\
 &&+\gamma\left(\omega_0-\Omega,t\right)
  \left(\frac{1}{2}\left|E_0\right>\left<E_{1,-}\right|\rho(t)\left|E_{1,-}\right>\left<E_0\right|
  -\frac{1}{4}\left\{\left|E_{1,-}\right>\left<E_{1,-}\right|,\rho(t)\right\}\right),
\end{eqnarray}
where
$\left|E_{1,\pm}\right>=(\left|1,g\right>\pm\left|0,e\right>)/\sqrt{2}$
are the eigenstates of $H_{JC}$ with one total excitation, with
energy $\omega_0/2\pm\Omega$, and
$\left|E_0\right>=\left|0,g\right>$ is the ground state, with
energy $-\omega_0/2$. The time-dependent decay rates for
$\left|E_{1,-}\right>$ and $\left|E_{1,+}\right>$ are
$\gamma(\omega_0-\Omega,t)$ and $\gamma(\omega_0+\Omega,t)$
respectively. Equation (\ref{masterequation}) is the non-Markovian
extension of the Markovian model introduced, for the same case, in
Ref.~\refcite{Scala}.

By direct substitution, it is possible to show that, if the system
starts from the state $\left|0,e\right>$, i.e., if the atom is
initially excited and the cavity is initially empty, the solution
of Eq.~(\ref{masterequation}) is the following one:
\begin{eqnarray}\label{rhorabinonmarkovian}
 &&\rho(t)=\left(1-\frac{1}{2}\mathrm{e}^{-\frac{I_-(t)}{2}}-\frac{1}{2}\mathrm{e}^{-\frac{I_+(t)}{2}}\right)
 \left|E_0\right>\left<E_0\right|+\frac{1}{2}\mathrm{e}^{-\frac{I_-(t)}{2}}\left|E_{1,-}\right>\left<E_{1,-}\right|\nonumber\\
 \nonumber\\
 &&+\frac{1}{2}\mathrm{e}^{-\frac{I_+(t)}{2}}\left|E_{1,+}\right>\left<E_{1,+}\right|-\frac{1}{2}\mathrm{e}^{-\frac{I_-(t)+I_+(t)}{4}}\left(\mathrm{e}^{2i\Omega t}\left|E_{1,-}\right>\left<E_{1,+}\right|+
\mbox{h.c.}\right),
\end{eqnarray}
where $I_\pm(t)=\int_0^t\gamma(\omega_0\pm\Omega,t')dt'$. From
Eq.~(\ref{rhorabinonmarkovian}) one can compute all the
populations we are going to show in the following.  Below we will
study the behavior of the non-Markovian time-dependent rates
$\gamma(\omega_0\pm\Omega,t)$, which, through the quantities
$I_\pm(t)$, lead to non-Markovian behavior and to population
trapping.

\section{The non-Markovian decay rates for a Lorentzian spectrum}
\label{ch5_secrates} As a first model of environment at zero
temperature with non-flat spectrum, we consider the Lorentzian
distribution~\cite{petruccionebook}:
\begin{equation}\label{lorentziandensity}
  J(\omega)=\frac{1}{2\pi}\frac{\alpha\lambda^2}{(\omega_1-\omega)^2+\lambda^2},
\end{equation}
where $\alpha$ is the system-environment coupling strength, and
$\lambda$ is the width of the distribution, describing also the
inverse of the reservoir memory time. The case of Lorentzian
spectrum is analytically treatable, while capturing important
features of the non-Markovian dynamics we are interested in, i.e.,
the time-dependence of the decay rates and their different
stationary values. We consider the case in which the spectrum is
peaked on the frequency of the state $\left|E_{1,-}\right>$, i.e.,
$\omega_1=\omega_0-\Omega$, where $\omega_0$ is the atomic Bohr
frequency and $\Omega$ is the Rabi splitting due to the JC
interaction.

The rate $\gamma(\omega,t)$ for a generic transition with Bohr
frequency $\omega$ is equal to
$\gamma(\omega,t)=2\mathrm{Re}\left\{\Gamma(\omega,t)\right\}$,
where $ \Gamma(\omega,t)$ is related to the spectral density
$J(\omega)$ through the relation\cite{Scala3}:
\begin{equation}\label{Gamma_t_dep}
 \Gamma(\omega,t)=\int_0^td\tau\int_{-\infty}^{+\infty}d\omega'\mathrm{e}^{i(\omega-\omega')\tau}J(\omega').
\end{equation}
This relation gives the right Markovian decay rates and Lamb
shifts for $t\rightarrow+\infty$.

By performing first the integral with respect to $\tau$, one
obtains:
\begin{eqnarray}\label{integraltau}
 \int_0^td\tau\mathrm{e}^{i(\omega-\omega')\tau}=\frac{\sin(\omega'-\omega)t}{\omega'-\omega}-i\frac{1-\cos(\omega'-\omega)t}{\omega'-\omega},
\end{eqnarray}
which, substituted into Eq. (\ref{Gamma_t_dep}), gives:
\begin{eqnarray}\label{Gammatdep_generalJ}
 \Gamma(\omega,t)=\int_{-\infty}^{+\infty}d\omega'J(\omega')\frac{\sin(\omega'-\omega)t}{\omega'-\omega}-\!
 i\!\!\int_{-\infty}^{+\infty}d\omega'J(\omega')\frac{1-\cos(\omega'-\omega)t}{\omega'-\omega}.
\end{eqnarray}
The second term in Eq. (\ref{Gammatdep_generalJ}) gives a
time-dependent Lamb shift which in the following will be
neglected. The first term, i.e., the real part, is half the
time-dependent decay rate $\gamma(\omega,t)$.

Specializing to the Lorentzian spectral density given in Eq.
(\ref{lorentziandensity}), the real part of Eq.
(\ref{Gammatdep_generalJ}) becomes:
\begin{eqnarray}\label{ReGammaLorentz}
 \mathrm{Re}\left\{\Gamma(\omega,t)\right\}
 %&=&\frac{1}{2\pi}\int_{-\infty}^{+\infty}d\omega'
 %\frac{\alpha\lambda^2}{(\omega_1-\omega')^2+\lambda^2}\frac{\sin(\omega'-\omega)t}{\omega'-\omega}\nonumber\\
 %\nonumber\\
 %&=&
 =\frac{1}{2\pi}\mathrm{Im}\left\{\mathrm{e}^{-i\omega t}\!\int_{-\infty}^{+\infty}\!d\omega'\frac{\alpha\lambda^2
 \mathrm{e}^{i\omega' t}}{[(\omega_1-\omega')^2+\lambda^2](\omega'-\omega)}\right\},
\end{eqnarray}
whose last form is suitable for an evaluation by means of the
method of the residues, by closing the integration path with a
circle of ray $R$ on the upper complex half-plane, where the
integral vanishes when $R\rightarrow\infty$\cite{dennery}. Such an
evaluation is straightforward. It is only worth noting that the
pole on the real axis, at $\omega'=\omega$, must be circumvented
from below, so that its residue is taken for half its value, with
positive sign. The reason of this choice is that, as we will see,
it gives a positive stationary value for the decay rate, while the
other choice would give a negative stationary decay rate, which
would be unphysical.

By taking twice the quantity in Eq. (\ref{ReGammaLorentz}) and
evaluating the integral, we finally obtain the following
expression for the decay rate $\gamma(\omega,t)$:
\begin{eqnarray}\label{gamma_tdep}
 &&\gamma(\omega,t)=\frac{\alpha\lambda^2}{(\omega_1-\omega)^2+\lambda^2}\left\{1+\left[\frac{\omega_1-\omega}{\lambda}\sin(\omega_1-\omega)t-
 \cos(\omega_1-\omega)t\right]\mathrm{e}^{-\lambda t}\right\}.
\end{eqnarray}

From Eq.~(\ref{gamma_tdep}) we clearly see the general behavior of
the time-dependent rates: all the rates $\gamma(\omega,t)$ are
zero at $t=0$, then they oscillate in time with a time-dependent
average value, till they reach stationary values for
$t\gg\lambda^{-1}$. These stationary values are proportional to
$J(\omega)$,~i.e.,~they are equal to the rates one obtains from a
Markovian theory. For these reasons, as anticipated, the quantity
$\lambda^{-1}$ can be seen as the memory time of the
system-reservoir interaction and non-Markovian effects are
expected to occur for times shorter than $\lambda^{-1}$.

In particular, taking the peak of the spectrum in
$\omega_1=\omega_0-\Omega$ and substituting
$\omega=\omega_0\pm\Omega$, one obtains the decay rates for the
two dressed states $\left|E_{1,\pm}\right>$, which, along with the
solution of the master equation in Eq.
(\ref{rhorabinonmarkovian}), allow us to compute the dynamics of
the atom cavity system presented in Ref.~\refcite{Scala3}, to
which we refer for the plots of the populations considered. The
population of the ground state of the atom-cavity system
$\left|0,g\right>$ increases quadratically for short times, with
some oscillations superimposed which are signatures of the
oscillations of the decay rates. For larger times, i.e.,
$t\gg\lambda^{-1}$, it increases in time as the sum of the two
exponentials, with rates $\gamma(\omega_0-\Omega, \infty)$ and
$\gamma(\omega_0+\Omega, \infty)$ respectively. It is easy to see
that the smaller the value of $\lambda$ in the spectrum, the
smaller the stationary decay rate $\gamma(\omega_0+\Omega,
\infty)$, while $\gamma(\omega_0-\Omega, \infty)=\alpha$ does not
change with $\lambda$. A consequence of this point is the
possibility of having situations where, starting with the atom
initially excited and the cavity with zero photons, while the
state $\left|E_{1,-}\right>$ has completely decayed, the
population of $\left|E_{1,+}\right>$ keeps a constant value for
long times, namely the ones in the interval
$\gamma(\omega_0-\Omega, \infty)^{-1}\ll
t\ll\gamma(\omega_0+\Omega, \infty)$. A consequence of this fact
is that the population of the atom is trapped in the excited state
for an amount close to $25\%$.

One may wonder how strongly this effect could depend on the
particular model of spectrum we have chosen. In fact, the
possibility of having a certain amount of population trapping in
the system depends on the stationary decay rates only, so that the
effect of population trapping involves only the values of the
spectrum at the Bohr frequencies of the transitions of interest,
and not the shape of the spectrum over all the real axis. To
clarify this point, in the next section we show the behavior of
the same populations when the spectrum of the environment has a
shape different from the one considered in this section.

\section{Evolution of the system in the case of a structured spectrum}
\label{sec_evolpop} As a second model for a non-flat spectrum, we
take a simple model of structured reservoir, consisting in a
Lorentzian background with a Lorentzian
gap\cite{petruccionebook,garraway2}:
\begin{equation}\label{garrawaygap}
  J(\omega)=\frac{1}{2\pi}\frac{\alpha_1\lambda_1^2}{(\omega_1-\omega)^2+\lambda_1^2}
  -\frac{1}{2\pi}\frac{\alpha_2\lambda_2^2}{(\omega_2-\omega)^2+\lambda_2^2},
\end{equation}
where the gap is given by the inverted Lorentzian peak at
frequency $\omega=\omega_2$. From the calculation of the previous
section, it is straightforward to show that each Lorentzian peak
gives a similar contribution, but with opposite sign, so that the
time-dependent decay rate of a transition of Bohr frequency
$\omega$ is given by:
\begin{eqnarray}\label{gamma_tdep_gap}
 &&\gamma(\omega,t)=\frac{\alpha_1\lambda_1^2}{(\omega_1-\omega)^2+\lambda_1^2}\left\{1+\left[\frac{\omega_1-\omega}{\lambda_1}\sin(\omega_1-\omega)t-
 \cos(\omega_1-\omega)t\right]\mathrm{e}^{-\lambda_1
 t}\right\}\nonumber\\
 \nonumber\\
 &&-\frac{\alpha_2\lambda_2^2}{(\omega_2-\omega)^2+\lambda_2^2}\left\{1+\left[\frac{\omega_2-\omega}{\lambda_2}\sin(\omega_2-\omega)t-
 \cos(\omega_2-\omega)t\right]\mathrm{e}^{-\lambda_2 t}\right\}.
\end{eqnarray}

In the following we will take $\omega_1=\omega_2=\omega_0+\Omega$,
and $\alpha_1>\alpha_2$ and $\lambda_1>\lambda_2$: in this way we
assume that the Bohr frequency corresponding to the transition
$\left|E_{1,+}\right>\rightarrow\left|E_{0}\right>$ corresponds to
the minimum of the spectral density, in analogy with what done in
the case of a single Lorentzian peak. The choice of
$\alpha_1=0.1*2\Omega$, $\alpha_2=0.099*2\Omega$,
$\lambda_1=100*2\Omega$ and $\lambda_2=0.1*2\Omega$ gives a ratio
$1/100$ between the stationary values of the two decay rates of
interest $\gamma(\omega_0\pm\Omega,\infty)$, a situation which is
close to the ideal case of perfect population trapping.

\begin{figure*}
 \begin{center}
  \begin{tabular}{rl@{\qquad}rl}
  \mbox{\footnotesize (a)}& \includegraphics[height=0.26\textwidth, angle=0]{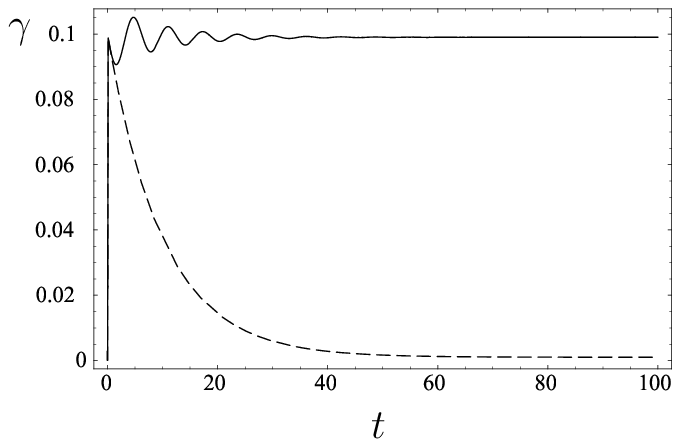}
  & \mbox{\footnotesize (b)}& \includegraphics[height=0.26\textwidth, angle=0]{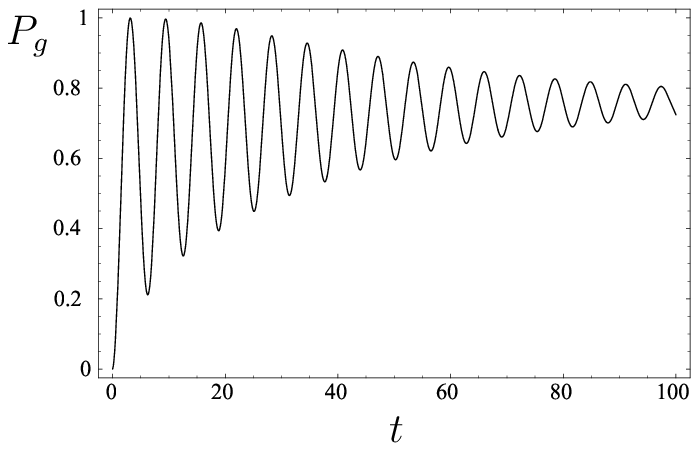}
  \end{tabular}
 \end{center}
 \caption{(a) Time-dependent decay rates of
    $\left|E_{1,-}\right>$ (solid line) and $\left|E_{1,+}\right>$
    (dashed line)
    as a function of $t$ (in units of $(2\Omega)^{-1}$). (b) Population of the atomic ground state as a function of $t$
    (in units of $(2\Omega)^{-1}$).}\label{figure}
\end{figure*}

As we see from Fig. 1-(a), the short-time behavior of the rates is
rather different from the case of the single Lorentzian peak,
indeed in the case of the gap the decay rate of
$\left|E_{1,+}\right>$ initially increases in the same way the
rate of $\left|E_{1,-}\right>$ does, and, after reaching its
maximum value close to $0.1*2\Omega$, it decays exponentially to
its stationary value $0.01*2\Omega$. Anyway this difference in the
behavior of the rates does not lead to observable effects in the
dynamics of the atomic populations. Indeed, comparing the
population of the atomic ground state in Fig. 1-(b) with the
corresponding population in Ref.~\refcite{Scala3}, we see that the
behavior predicted by both the single-peak Lorentzian model and
the Lorentzian gap model is exactly the same: for a long time a
population trapping occurs in the excited atomic state, for an
amount of about $25\%$, no matter which of the two models one
chooses.

Summarizing, in the situation we have analyzed, the essential
point to properly choose the model is to check the appropriate
ratio between the stationary values of the decay rates. In this
sense, the most important property is the value of the spectrum at
the Bohr frequencies of the transitions of interest. On the other
hand, the particular shape of the spectrum one chooses affects
only the details of the short-time dynamics, especially the
short-time behavior of the decay rates, but the main aspects of
the dynamics of the atom-cavity system are not affected by its
choice.

We feel that these conclusions are quite general and that they
apply to a wide variety of situations involving lossy cavities
with non-flat spectra, as in the most advanced quantum
electrodynamic systems\cite{lambropoulosreview}.

\section*{Acknowledgements}
M.S. thanks the Fondazione Angelo Della Riccia for financial
support. S.M., J.P. and K.-A.S. acknowledge financial support from
the Academy of Finland (Projects No.~108699, No.~115682, and
No.~115982),
%(projects 108699, 115982, 115682)
the Magnus Ehrnrooth Foundation and the V\"{a}is\"{a}l\"{a}
Foundation. A.M. acknowledges partial support by MIUR project
II04C0E3F3
\textit{Collaborazioni Interuniversitarie ed Internazionali Tipologia C}.
%\textit{COLLABORAZIONI INTERUNIVERSITARIE ED INTERNAZIONALI TIPOLOGIA C}.

\end{document}